\documentclass[twocolumn,amssymb,amsmath,nobibnotes,nofootinbib,showpacs,aps,prl]{revtex4}
\usepackage[dvips]{graphicx,color}

\usepackage{hyperref}
\hypersetup{
    colorlinks=true,
    linkcolor=blue,
    citecolor=blue,
    urlcolor=blue,
    bookmarksopen=true,
    bookmarksnumbered=true,
    pdftitle={SBT Letter}
}

\begin{document}

\title{Unconventional topological phase transition from semimetal to insulator in SnBi$_2$Te$_4$: Role of anomalous thermal expansion}
\author{T. K. Dalui$^1$}
\author{B. Das$^2$}
\author{C. K. Barman$^3$}
\author{P. K. Ghose$^1$} 
\author{A. Sarma$^4$} 
\author{S. K. Mahatha$^{5}$}
\author{F. Diekmann$^6$} 
\author{K. Rossnagel$^{6,7}$}
\author{S. Majumdar$^1$} 
\author{A. Alam$^2$}
\email{aftab@phy.iitb.ac.in} 
\author{S. Giri$^1$}
\email{sspsg2@iacs.res.in}
\affiliation{$^1$School of Physical Sciences, Indian Association for the Cultivation of Science, Jadavpur, Kolkata 700032, India \\
$^2$Department of Physics, Indian Institute of Technology Bombay, Mumbai 400076, India \\
$^3$Department of Physics, Sungkyunkwan University, Suwon 16419, Republic of Korea \\
$^4$Deutsches Elektronen-Synchrotron DESY, 22607 Hamburg, Germany \\
$^5$UGC-DAE Consortium for Scientific Research, Khandwa Road, Indore 452001, India \\
$^6$Institute of Experimental and Applied Physics, Kiel University, 24098, Germany \\
$^7$Ruprecht Haensel Laboratory, Deutsches Elektronen-Synchrotron DESY, 22607 Hamburg, Germany}

\begin{abstract}
We propose SnBi$_2$Te$_4$ to be a novel candidate material exhibiting temperature ($T$) mediated transitions between rich topological phases. From a combined theoretical and experimental studies, we find that SnBi$_2$Te$_4$ goes from a low-$T$ topological semimetallic phase to a high-$T$ (room temperature) topological insulating phase via an intermediate topological metallic phase. Single crystals of SnBi$_2$Te$_4$ are characterized by various experimental probes including Synchrotron based X-ray diffraction, magnetoresistance, Hall effect, Seebeck coefficient, magnetization and angle-resolved photoemission spectroscopy (ARPES). X-ray diffraction data confirms an anomalous thermal expansion of the unit cell volume below $\sim$ 100 K, which significantly affects the bulk band structure and hence the transport properties, as confirmed by our density functional theory calculations.  Simulated surface states at 15 K agree fairly well with our ARPES data and are found to be robust with varying $T$. This indirectly supports the experimentally observed paramagnetic singularity in the entire $T$-range. The proposed coexistence of rich topological phases is a rare occurrrence, yet paves a fertile ground to tune various topological phases in a material driven by structural distortion.

\end{abstract}
\maketitle

Understanding the nature of topological states of matter beyond topological insulators (TI) is currently one of the prime focuses in condensed matter physics \cite{has10}. These states can be realized in metals or semimetals, which are conventionally termed as topological metals or semimetals and have  attracted significant attention \cite{you12,xu15,yan14,pot14}. In few bulk metals or semimetals adjacent to TIs, such as Sb \cite{hsi09,hsixiv}, Bi$_{0.91}$Sb$_{0.09}$ \cite{tas09,tas10}, Bi$_{2-x}$Mn$_x$Te$_3$ \cite{hor10}, and Bi$_2$Se$_3$ \cite{ana10,eto10}, which was initially proposed to be a wide band-gap TI, interesting topological surface states have been witnessed. The emergence of new topological metals or semimetals continues to grow in interest and opens up a new frontier in the field of quantum materials \cite{ran09,ber10,ber11,ort14}.

Bi$_2$Te$_3$ is one of the celebrated three-dimensional (3D) topological insulators \cite{chen09}. Recently, the tetradymitelike-layered pseudobinary compounds having a general chemical formula A$^{IV}$Te$-$Bi$_2$Te$_3$ (A$^{IV}$ = Ge, Sn, Pb) or (Ge,Sn,Pb)Bi$_2$Te$_4$ have attracted special attention because of their promising thermoelectric properties \cite{she04,zha10,sam20,zha22,kur13,pan15,lee18,gur21}. Out of the series, PbBi$_2$Te$_4$ has been experimentally verified as a 3D topological insulator with a considerably large bulk energy gap of 230 meV, which was further validated by {\it ab-initio} band structure calculations \cite{kur12}. In contrast, a topological metallic state was proposed for GeBi$_2$Te$_4$ from both experiment and density functional theory (DFT) calculations \cite{mar13}. Recently, a strong electron-phonon coupling has been verified in GeBi$_2$Te$_4$ based on  Raman studies, where the electrons of the topological state were proposed to interact with the phonons involving vibrations of Bi-Te bond \cite{sin22,li21}. Recent studies on epitaxial (SnBi$_2$Te$_4$)$_n$(Bi$_2$Te$_3$)$_m$ superlattices suggested an overlap between topological surface and bulk conduction band (BCB) states at the Fermi level. The contribution of BCB states was found to reduce by increasing the Sn content \cite{fra21}. Nevertheless, the topological phase of pristine SnBi$_2$Te$_4$ still needs to be settled, though the possibility of a TI state has been proposed theoretically \cite{vil16,zou18,yan12,ver15,men15}.

In this letter, we observe an unusual coexistence of topologically non-trivial electronic phases in SnBi$_2$Te$_4$ which is intimately related to negative thermal expansion with temperature. Our \textit{ab-initio} results match reasonably well with the surface band structure and Fermi surface obtained from ARPES measurements at 15 K. In accordance with the prediction of the DFT calculations, the paramagnetic singularity in the entire temperature range of 2$-$300 K points to a robust Dirac surface state. The Hall effect distinguishes the surface and bulk conduction mechanisms, which are also influenced by the anomalous thermal expansion, as confirmed by synchrotron based X-ray diffraction measurements. Though the temperature mediated structure distortion remains mild in most materials, its influence on the evolution of electronic structure can be significant which is rarely studied in the literature \cite{sun15,kun18}. Here, we will exploit this aspect and show how a structural distortion driven conduction mechanism can open a new route to design and tuning of topologically non-trivial electronic phases in SnBi$_2$Te$_4$.

\begin{figure}[t]
\centering
\includegraphics[width= \columnwidth]{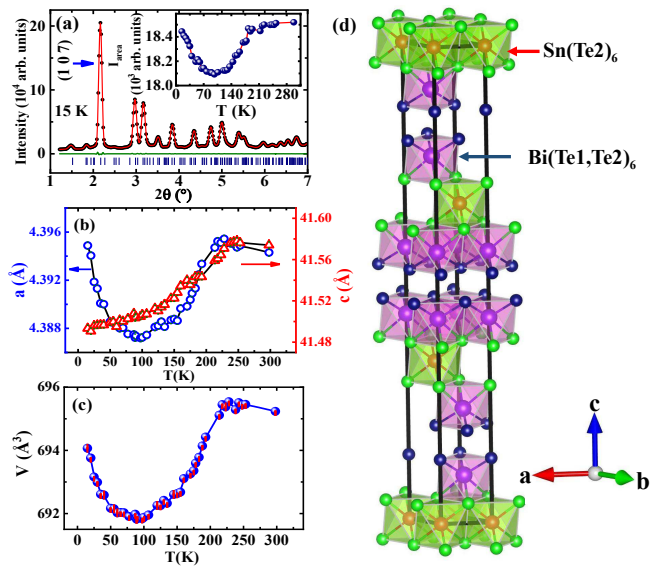}
\caption {(a) Rietveld refinement of the synchrotron diffraction pattern at 15 K. Inset shows $T$ variation of the integrated intersity ($I_{\rm area}$) of the (107) peak. $T$ variations of the (b) refined lattice constants (a and c) and (c) unit cell volume ($V$). (d) Crystal structure of SnBi$_2$Te$_4$ based on the refined data.}
\label{str}
\end{figure}

Sample preparation, experimental measurements and computational details can be found in the Supplemental Material (SM) \cite{sup}.

 Analysis of the synchrotron diffraction patterns using Rietveld refinement indicates that the same structure is retained in the entire recorded temperature ($T$) range. Structural refinement at a representative temperature (15 K) is depicted in Fig. \ref{str}(a). The Rietveld refined coordinates are Sn (0,0,0), Te1 (0,0,0.13685(1)), Te2 (0,0,0.26758(8)), and Bi (0,0,0.42963(7)) with the refinement reliability parameters, $R_w$ (\%) = 8.2, $R_{exp}$ (\%) = 4.15, and $\chi^2$ = 1.56. Inset of Fig. \ref{str}(a) depicts thermal variation of the integrated intensity (I$_{\rm area}$) of the (107) peak. The I$_{\rm area}$ decreases with $T$ almost linearly until $\sim$ 170 K, below which it further decreases rapidly, following a significant increase below $\sim$ 100 K down to lowest recorded temperature. The lattice constant ($a$) and unit cell volume ($V$) follow a similar $T$-dependence, see Fig. \ref{str}(b,c). The anomalous low-$T$ increase of $V$ is considerable as $\sim$ 0.33 \% at 15 K with respect to the value at 100 K. Figure \ref{str}(d) shows the stacking of Sn(Te2)$_6$ and Bi(Te1,Te2)$_6$ octahedra along the $c$-axis, as obtained from the refined atomic positions.

\begin{figure}[t]
\centering
\includegraphics[width =\columnwidth]{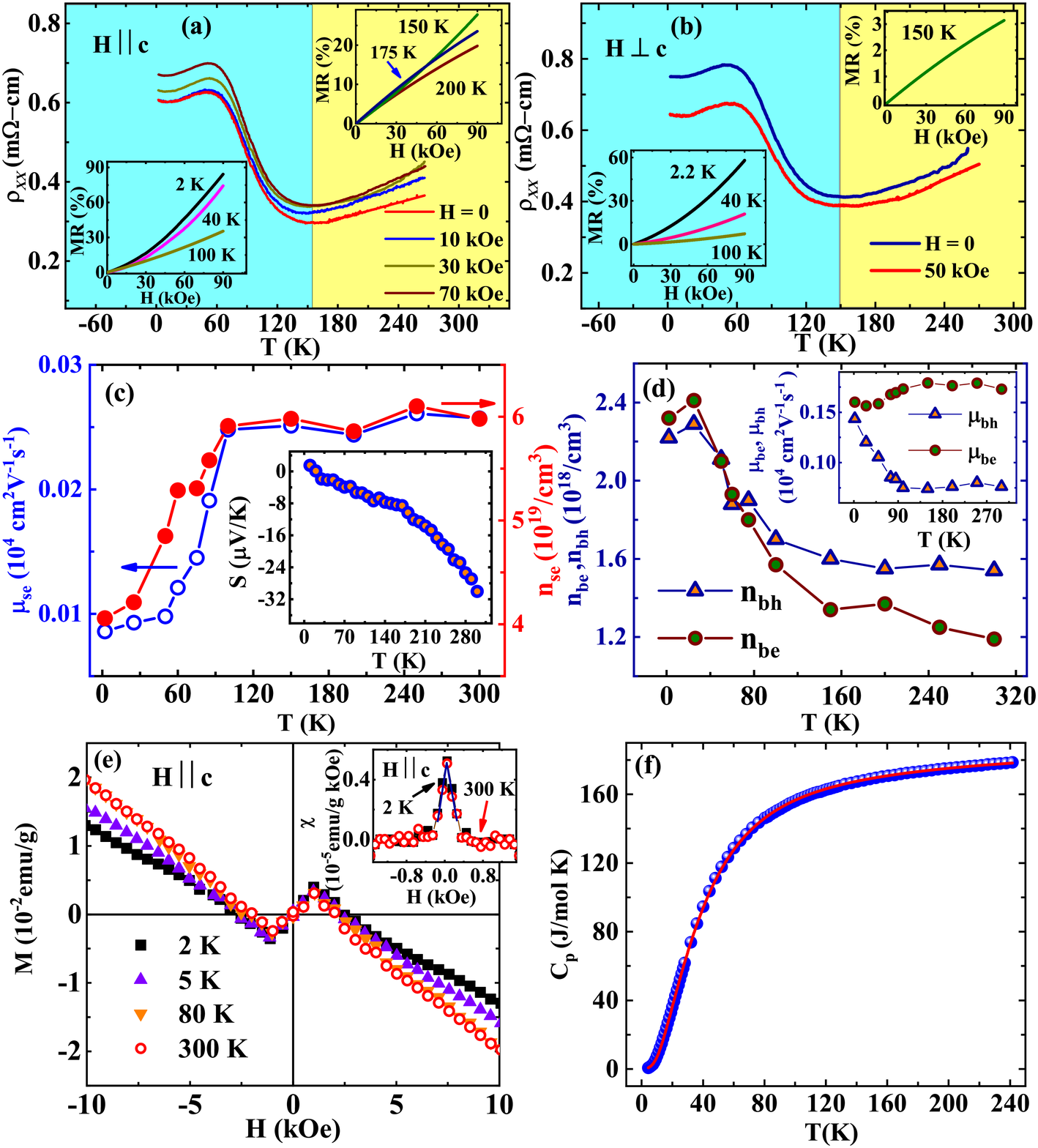}
\caption { Thermal ($T$) variations of $\rho_{xx}$ recorded at selected field ($H$) applied (a) $||$, (b) $\bot$ to the [001] direction. Lower (upper) insets of (a) and (b) show low-$T$ (high-$T$) magnetoresistance (MR) vs. H curves at selected T. T-dependence of (c) surface-electron mobility, $\mu_{se}$ (left axis) and surface-electron carrier density, $n_{se}$ (right axis), (d) bulk-electron carrier density, $n_{be}$, and bulk-hole carrier density, $n_{bh}$. Inset of (c) and (d) shows $T$ variations of Seebeck coefficient ($S$) and $\mu_{be}$, $\mu_{bh}$ respectively. (e) Magnetization ($M$) vs. $H$ at selected $T$. Inset shows $\chi$ vs. $H$ plot at 2 and 300 K. (f) Specific heat capacity ($C_p$) vs. $T$. Solid curve shows the fit, as described in the text.}
\label{MR}
\end{figure}

Figures \ref{MR}(a) and \ref{MR}(b) show a $T$ variation of resistivity ($\rho_{xx}$) with the magnetic field ($H$) applied $||$ and $\bot$ to the [001] direction, respectively. Here, the subscript `$xx$' represents same direction of the applied current and the recorded voltage, where the electrical connections are made on the (001) plane. For both the $||$ and $\bot$ components, the $\rho_{xx}(T)$ show typical metallic character down to $\sim$ 150 K, below which a semiconducting behavior is observed. A maximum in $\rho_{xx}(T)$ is noticed around $\sim$ 60 K, which is followed by a decreasing trend down to 2 K in both cases. The $\rho_{xx}(T)$ in the present system is similar to that observed in Bi$_2$Se$_3$ \cite{che09,but10}, Bi$_2$Te$_3$ \cite{kim12}, and Bi$_2$Te$_2$Se and Bi$_{2-x}$Sb$_x$Te$_2$S \cite{cai18}. $\rho_{xx}(T)$ for Bi$_2$Se$_3$ has been characterized by its dominant bulk contribution  \cite{che09,but10}. On the other hand, the antisite disorder in Bi$_2$Te$_3$ was suggested to be responsible for low-$T$ semiconducting behavior. For Bi$_2$Te$_2$Se and Bi$_{2-x}$Sb$_x$Te$_2$S, surface and bulk contribution to $\rho$ were proposed to be independent, and giving rise to a semiconducting-like behavior in low T regime dominant by surface states \cite{cai18}. For SnBi$_2$Te$_4$, the value of $\rho_{xx}$ lies in the range $\sim$ 0.3-0.8 m$\Omega$-cm, which is at least one order of magnitude lower than the reported results, pointing towards a more dominant metallic character in SnBi$_2$Te$_4$.

The magnetoresistance (MR), defined as [$\rho_{xx}(H)$ - $\rho_{xx}(H=0)$]/$\rho_{xx}(H=0)$, curves at selected $T$ are shown in the insets of Fig. \ref{MR}(a) and \ref{MR}(b) for the $||$ and $\bot$ components respectively, where lower and upper insets in both the panels display the low-$T$ ($<$ 150 K) and high-$T$ curves, respectively. As evident from the upper inset of Fig. \ref{MR}(a), nature of the curves at high-$H$ in the high-$T$ region ($> 150$ K) are different from the low-$T$ MR curves. The details of the MR curves are shown in Figs. S2(a) and S2(b) of SM \cite{sup}. The orbital contribution to MR in the nonmagnetic metals was explained by Kohler, known as the Kohler's rule (KL) \cite{koh38}. The semi-classical KL has been tested for various systems viz., quasi-2D metal \cite{mck98}, graphite and bismuth \cite{kop06}, Weyl semimetal \cite{wan15}, topological semimetals \cite{jo17,dal21}. Kohler's rule is verified for both the $||$ and $\bot$ components of MR, see Figs. S2(b) and S2(c) of SM \cite{sup}. The MR curves for the $||$ component follow a single master curve for $T <$ 15 K, above which they deviate and follow the KL until 150 K. The deviation from the master curve of KL was recently explained via a $T$-dependent change in carrier densities \cite{xu21}, which is followed in our present system as well, as explained below. The rule is satisfied for the $\bot$ component of MR as well, but does not follow a single master curve.

The mobilities and carrier densities are obtained from the $H$-dependent Hall resistivity ($\rho_{yx}$), as shown in Fig. S2(e) of SM \cite{sup}. Interestingly, conductivities ($\sigma$) do not fit a two-band model \cite{hua15,dal21}, rather a three-band model \cite{par18,ant96} is required to correctly capture all the features. The expression for three-band model and satisfactory global fits (see Fig. S2(f) of SM) of $\sigma$ are shown in SM \cite{sup}. The three-band model suggests the two electron-type Fermi pockets and one hole-type Fermi pocket, implying coexistence of surface and bulk conductivities. The bulk conductivity is governed by the electron ($\mu_{be}$) and hole ($\mu_{bh}$) mobilities. Let us denote $n_{se}$, $n_{be}$, and $n_{bh}$ as the carrier densities of surface and bulk electrons, and bulk holes, respectively, while $\mu_{se}$ as the mobility of the surface electrons. As depicted in Fig. \ref{MR}(c), $n_{se}$ and $\mu_{se}$ follow similar $T$-dependence i.e. nearly $T$-independent until $\sim$ 120 K, below which a sharp decrease is observed. Both parameters show a nearly $T$-independent behavior at low-$T$. The $n_{be}(T)$ and $n_{bh}(T)$ are nearly $T$-independent until $\sim$ 120 K (see Fig. \ref{MR}(d)), below which they increase with decreasing $T$ and indicate a nearly $T$-independent trend at low $T$. The $\mu_{be}$ and $\mu_{bh}$ are shown in the inset of Fig. \ref{MR}(d), which are nearly $T$-independent until $\sim$ 120 K, below which both of them show contrasting behavior. $\mu_{be}$ decreases, whereas $\mu_{bh}$ increases below $\sim$ 120 K. The Inset of Fig. \ref{MR}(c) displays the Seebeck coefficient ($S$) vs. $T$, exhibiting a non-linear behavior, indicating the existence of more than one type of charge carrier. The value of $S$ is $\sim$ -30 $\mu$V/K at 300 K, which indicates dominant electron conduction, consistent with the Hall measurement results.

Figure \ref{MR}(e) shows the magnetization curves ($M-H$) at selected $T$ between 2 and 300 K, where $H$ is applied $||$ to [001]. The $M-H$ curves show a diamagnetic character over a wide $H$-range except at low-$H$ ($\sim\pm$0.3 kOe), where a typical paramagnetic behavior is observed. The susceptibility ($\chi$) is obtained from the first-order derivative of the $M-H$ curve with respect to $H$. The height of the peak in $\chi$ over the diamagnetic background remains almost the same at 2 and 300 K, as evident in the inset. The paramagnetic singularity at low-$H$ has been proposed due to the helical spin texture of 2D Dirac fermions on the surface, as also suggested for (Bi,Sb)$_2$(Se,Te)$_3$ \cite{zha14}, ZrTe$_5$ \cite{par16}, and Bi$_{1.5}$Sb$_{0.5}$Te$_{1.7}$Se$_{1.3}$ \cite{dut17}.

To confirm the absence of any structural phase transition, $T$-variation of the specific heat capacity $C_{p}(T)$ was recorded in zero-field from 4 to 300 K, as shown in Fig. \ref{MR}(f). The $C_{p}(T)$ data does not show any anomaly over the recorded $T$-region. The $C_{p}(T)$ data in the intermediate temperature range does not fit well with either Debye or Einstein model. However, a trial with a combined Debye and Einstein model satisfactorily fits the experimental data using \cite{shang19,kumar22,chat20,ban18,das20}, $C_{p}(T) = pD(\theta_{D},T) + (1-p)E(\theta_{E},T)$. The combined Debye and Einstein model represent the acoustic and optical phonon-mode contributions \cite{tar03}. The best fit is shown by the solid curve in Fig. \ref{MR}(f) with $\theta_{D}$ = 105.7 K, $\theta_{E}$ = 175.1 K, and $p$ = 0.59. The low-$T$ linear temperature coefficient of $C_p$ is found to be 32.1 mJ mol$^{-1}$K$^{-2}$ (obtained by subtracting the lattice contribution) which is comparable to the reported results for a TI \cite{phe14}.

\begin{figure*}[t]
    \centering
    \includegraphics[width =2\columnwidth]{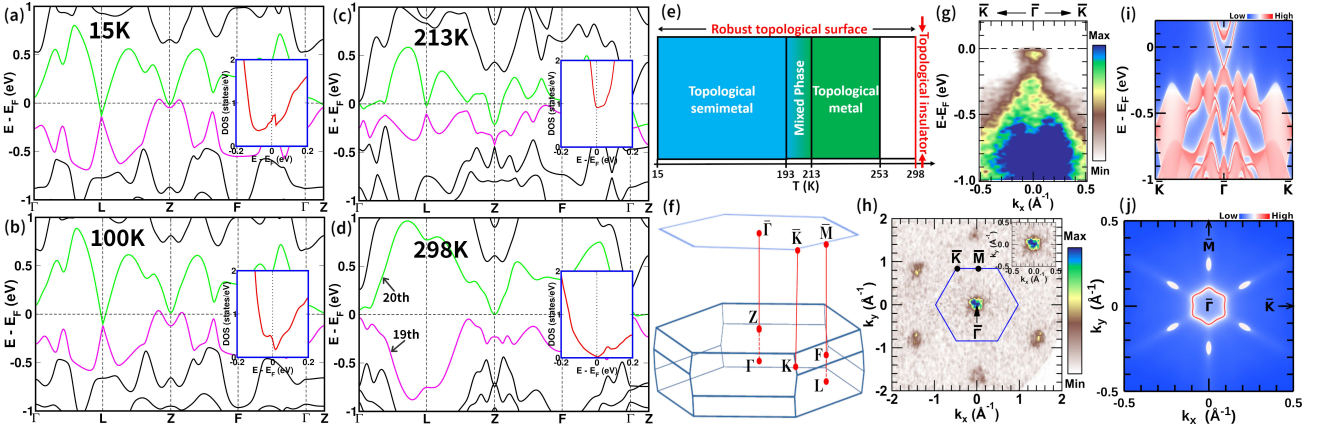}
    \caption {(Colour online) Bulk band structure of SnBi$_2$Te$_4$ using experimental lattice parameters at (a) 15 K, (b) 100 K, (c) 213 K, (d) 298 K with the corresponding total DOS shown in the insets and the 19th and 20th bands highlighted. (e) Schematic of $T$-dependent electronic phase diagram. (f) Bulk and (111) surface-projected Brillouin zone (BZ) (g,h) ARPES bands and Fermi surface (FS) at 15 K with the first BZ highlighted in blue. The inset shows a zoomed in FS around $\overline{\Gamma}$.  (i,j) Simulated surface bands and FS using relaxed internal coordinates of 15 K structure.}
    \label{cal}
\end{figure*}

To better understand our experimental data, electronic band structures and density of states (DOS) of SnBi$_2$Te$_4$ were simulated using the experimental lattice parameters and Wyckoff positions (henceforth referred as ELEW) obtained at different $T$. Computational details are given in SM \cite{sup}. The results at a few representative $T$ (15 K, 100 K, 213 K, 298 K) are shown in Fig. \ref{cal}(a-d) with the DOS near E$_{\text{F}}$ shown in insets. For electronic structures at few other intermediate $T$, we refer the reader to Fig. S3 of SM \cite{sup}.  At 15 K, one can notice the presence of a relatively moderate DOS with a small peak at E$_{\text{F}}$ (see inset of Fig. \ref{cal}(a)). This small peak remains upto $\sim$60 K which indirectly supports the increasing and hence metallic trend of $\rho_{xx}$ in the range $\sim$2$-$60 K. Beyond 60 K, this peak begins to shift below E$_{\text{F}}$ and slowly disappears (see inset of Fig. \ref{cal}(b)). This again corroborates with the decreasing, and hence semiconducting trend of $\rho_{xx}$ in the range $\sim$60$-$150 K. At 213 K, it shows a relatively high DOS near E$_{\text{F}}$  (see Fig. \ref{cal}(c) inset), explaining an upturn and hence a weak metallic trend of $\rho_{xx}$ in the range $\sim$150$-$300 K, with a lower magnitude of $\rho_{xx}$ than that seen in the $\sim$2$-$60 K range. Interestingly, at 298 K, there is almost vanishing DOS at/near E$_{\text{F}}$ indicating a semiconducting bulk nature. Thus, the electronic DOS gives us a hint at a possible competition between semiconducting/semimetallic and metallic phases in our system. The origin of such diverse electronic phases is attributed to the sensitive changes in the internal atomic coordinates (Wyckoff positions) with $T$. The lattice parameters of SnBi$_2$Te$_4$ at both 15 K and 298 K are very close to each other (within $\sim$0.2\%) but their Wyckoff positions are very different. This leads to very different bulk band structures mediated by the impact on the bond lengths which in turn affects the hybridization between different atoms. This change in atomic coordinates does not stimulate any structural transition or symmetry-breaking. The band structures are especially sensitive to the position of Te2 atoms which are sandwiched between Sn and Bi atoms and hence helps in the formation of metallic octahedra around both Sn and Bi. Such sensitive dependence on ELEW has also been explored earlier in a different class of topological materials belonging to the same space group \cite{lin10,kuroda10,chang11,eremeev11,ohtsubo}.

We further performed parity analysis of the filled bands to compute the $\mathbb{Z}_2$ invariants \cite{fu2007} at the representative $T$. Other than 298 K, the electrons were partially filled up to the 20th band, while at 298 K, the 20th band shifts above E$_{\text{F}}$ and the 19th band becomes the topmost filled band. From parity analysis, we obtain $\mathbb{Z}_2 (\nu_0) = 1$ for SnBi$_2$Te$_4$ at all $T$ with the change in parity occurring either at $\Gamma$ or Z point in the bulk (see Fig. S4 of SM \cite{sup}). Hence, SnBi$_2$Te$_4$ is a strong TI, as reported earlier. Correspondingly, band inversion between Bi-$p$ and Te-$p$ orbitals is prominently visible along the $\Gamma-$Z direction at all $T$ (see Fig. S3 of SM). This guarantees the presence of topologically non-trivial surface states (TSS) at all $T$ due to the bulk-boundary correspondence. This is a remarkable result since it shows the robustness of the TSS and their insensitivity to $T$ in spite of the complicated hybridizations reflected in the bulk bands near E$_{\text{F}}$ in the entire $T$-range. Hence, we propose a schematic electronic phase diagram for SnBi$_2$Te$_4$, shown in Fig. \ref{cal}(e). Due to scarce experimental structural data, the range 193 K $<T<$ 213 K is labelled `Mixed Phase'. The range 253 K $<T<$ 298 K is kept blank due to no experimental data. Further studies on SnBi$_2$Te$_4$ at different $T$ may help identify the nature of the crossover regions and whether these changes in the electronic structure are abrupt or gradual in nature.

To probe the surface states experimentally, ARPES measurements were done at 15 K. The results are shown in Figs. \ref{cal}(g) and \ref{cal}(h). A conical outline is clearly visible in Fig. \ref{cal}(g) confirming the presence of a Dirac cone with the nodal point at -0.16 $\pm$ 0.01 eV below E$_{\text{F}}$. The Fermi surface shown in Fig. \ref{cal}(h) shows the isolated nature of the surface states with high local DOS at/around $\overline{\Gamma}$ only. In ternary layered materials such as SnBi$_2$Te$_4$, surface relaxation plays an important role and can have significant effects on the surface band structure \cite{chang11,eremeev11}. Simulated surface band structure and Fermi surface of SnBi$_2$Te$_4$ using relaxed internal atomic coordinates are shown in Figs. \ref{cal}(i) and \ref{cal}(j). Simulations using relaxed coordinates at other $T$ yield similar results. The expected Dirac-like dispersion is seen around $\overline{\Gamma}$ with the nodal point at $\sim$ -0.156 eV below E$_{\text{F}}$. The corresponding Fermi surface shows the well-known hexagonal warping of the Dirac cone away from the nodal point energy. There are 6 symmetric hole pockets around the cone originating from trivial surface states crossing E$_{\text{F}}$ along the $\overline{\Gamma}-\overline{\text{M}}$ direction. The corresponding helical spin textures are shown in Fig. S5 of SM \cite{sup}. The absence of hole pockets in ARPES data may be attributed to the use of soft X-rays which was not sufficient to resolve such finer details. For completeness, we have also simulated the surface dispersion using ELEWs at a few representative $T$, as shown in Fig. S5 of SM \cite{sup}. Helical TSS is observed in all of them which further highlights their robustness and the surface sensitivity, as discussed earlier. The above results also correspond well with the experimental signature of paramagnetic singularity obtained at both 2 K and 300 K (see inset of Fig. \ref{MR}(c)) as mentioned earlier.

It is worth noting that the overall magnitude of $\rho_{xx}$ is of the order of $m\Omega$, which suggests dominant metallic type transport. This can now be attributed to the TSS since $n_{se}$ is also an order of magnitude higher than $n_{be}$ and $n_{bh}$ (see Fig. \ref{MR}(c) and \ref{MR}(d)). Thus, one can assert that the overall transport is metal-like due to the TSS but the finer changes in $\rho_{xx}$ with $T$ have a bulk origin. The Hall conductivities (see Fig. S2 of SM \cite{sup}) were fitted with a three-band model, indicating the presence of three types of charge carriers, which further aligns with the outcome of our simulated electronic structure analysis. Semimetals have both electrons and holes as carriers in the bulk. The presence of TSS below E$_{\text{F}}$ provides electrons as carriers on the surface.

In summary, we present unconventional temperature mediated topological electronic phase transitions in SnBi$_2$Te$_4$. The system goes from a low-$T$ topological semimetallic phase to a high-$T$ (room temperature) topological insulator phase via an intermediate topological metal phase. Such unusual transition is intimately connected to the temperature dependent experimental lattice parameters, as demonstrated by our synchrotron based XRD data. The inherent nature of these phases are distinguished by \textit{ab-initio} calculations done with the $T$-dependent experimental lattice parameters. Our simulated results at 15 K match fairly well with transport and ARPES data. Magnetization measurements confirm the presence of a paramagnetic singularity in the entire $T$-range (2$-$300 K), which supports the robustness of topological surface states obtained from the simulations. The coexistence of such unconventional electronic phases in a single compound is not common and hence gives us a new avenue to explore the evolution of such phases via $T$-mediated structural changes. References \cite{zhang,wagner,book1,sin,hurd,kuznet,dal20_see,ton66,hoh64,kre96,kre96r,kre93,blo94,kre99,per96,grimme10,mar97,sou01,mar12,mos08,piz20,lee81,lee81r,san85,wu18} have been cited in the SM \cite{sup}.

\vspace{0.2in}
\noindent
{\bf Acknowledgements:}
Synchrotron based X-ray diffraction and photoemission studies were performed at the light source PETRA III of DESY, a member of the Helmholtz Association (HGF). Financial support (Proposal No. I-20200322 \& I-20191102) by the Department of Science \& Technology (Government of India) provided within the framework of the India@DESY collaboration is gratefully acknowledged. AA acknowledges DST- SERB (Grant No. CRG/2019/002050) for funding to support this research.

\end{document}